\documentclass[5p,times]{elsarticle}

\usepackage{caption}
\usepackage{subcaption}
\usepackage{multirow}
\usepackage{makecell}
\usepackage{hhline}
\usepackage{hyperref}
\usepackage{natbib}

\usepackage{xcolor}

\usepackage{xcolor}
\usepackage{soul,etoolbox}
\setstcolor{blue}
\usepackage{relsize}
\usepackage{mathptmx}
\newtoggle{finalPaper}
\toggletrue{finalPaper} 

\iftoggle{finalPaper} {
    \newcommand{\addtxt}[1]{#1}
    \newcommand{\change}[2]{#2}
    \newcommand{\rmvtxt}[1]{}
} {
    \newcommand{\addtxt}[1]{\textcolor{blue}{#1}}
    \newcommand{\change}[2]{\st{#1}\textcolor{blue}{#2}}
    \newcommand{\rmvtxt}[1]{\st{#1}}
}

\journal{Future Generation Computer Systems}

\begin{document}
\begin{frontmatter}

\title{A Big Data Architecture for Early Identification and Categorization of Dark Web Sites}

\author[addr1]{Javier Pastor-Galindo\corref{cor1}}
\ead{javierpg@um.es}

\author[addr2]{Hông-Ân Sandlin}
\ead{hongan.sandlin@ar.admin.ch}

\author[addr1]{F\'elix G\'omez M\'armol}
\ead{felixgm@um.es}

\author[addr2]{Gérôme Bovet}
\ead{gerome.bovet@ar.admin.ch}

\author[addr1]{Gregorio Mart\'inez P\'erez}
\ead{gregorio@um.es}

\cortext[cor1]{Corresponding author}
\address[addr1]{Department of Information and Communications Engineering, University of Murcia, 30100 Murcia, Spain}
\address[addr2]{Cyber-Defence campus, armasuisse Science and Technology, Switzerland}

\begin{abstract}
The dark web has become notorious for its association with illicit activities and there is a growing need for systems to automate the monitoring of this space. This paper proposes an end-to-end scalable architecture for the early identification of new Tor sites and the daily analysis of their content. The solution is built using an Open Source Big Data stack for data serving with Kubernetes, Kafka, Kubeflow, and MinIO, continuously discovering onion addresses in different sources (threat intelligence, code repositories, web-Tor gateways, and Tor repositories), downloading the HTML from Tor and deduplicating the content using MinHash LSH, and categorizing with the BERTopic modeling (SBERT embedding, UMAP dimensionality reduction,  HDBSCAN document clustering and c-TF-IDF topic keywords). In 93 days, the system identified 80,049 onion services and characterized 90\% of them, addressing the challenge of Tor volatility. A disproportionate amount of repeated content is found, with only 6.1\% unique sites. From the HTML files of the dark sites, 31 different low-topics are extracted, manually labeled, and grouped into 11 high-level topics. The five most popular included sexual and violent content, repositories, search engines, carding, cryptocurrencies, and marketplaces. During the experiments, we identified 14 sites with 13,946 clones that shared a suspiciously similar mirroring rate per day, suggesting an extensive common phishing network. Among the related works, this study is the most representative characterization of onion services based on topics to date.
\end{abstract}


\end{frontmatter}

\section{Introduction}
The dark web is the anonymous portion of the Internet, not indexed by well-known search engines, that has gained notoriety due to its association with illegal activities such as drug trafficking, human trafficking, and cybercrime~\cite{8954668}. Recently, this space has been exploited in the Ukrainian-Russian conflict as a platform for data leaks, propaganda dissemination, cyberwarfare, illicit arms trading, organized activism or censorship avoidance, highlighting the importance of the cyberdimension in nowadays problems~\cite{willett2022cyber}.

In particular, Tor (The Onion Routing) is the most popular dark web network used for anonymous communication and web browsing~\cite{jcp1030025}, which is the focus of this paper. Given the illicit nature of Tor, there is a growing need for automatic systems and digital platforms to automate the monitoring of this space to contribute to situational awareness, generate intelligence, support decision-making, identify emerging threats or perform a quick reaction \cite{ruiz2023general}. \addtxt{Manual search and individual categorization prove impractical, necessitating technological assistance for online identification and topic extraction. This is particularly crucial for early and scalable detection of illicit services, assisting investigators and law enforcement agencies.}

Several crawlers and platforms in the literature collect and analyze Tor onion services. However, they employ traditional schemes and outdated techniques, which are not efficient and scalable for continuous and intensive monitoring \cite{PASTOR}. \change{Additionally, sophisticated solutions are needed to face several challenges of volatility, redundancy, and variety related to the nature of Tor sites}{This work addresses these weaknesses and contributes to the state-of-the-art by proposing an innovative solution with modern techniques for effective dark web monitoring, facing the challenges of volatily, redundancy and variety related to the nature of Tor sites}.

Firstly, earlier work has found that after 24 hours of publication, fewer than half of the observed onions are reachable~\cite{11}. Highly volatile and short-lived services impact crawling performance, in which the number of onion addresses found is much higher than the number of onion addresses that are still reachable for analysis afterwards~\cite{43}. Most discovered links are already offline by the time they are published in related work, having a huge discrepancy between identified services and active ones due to the time between deployment, identification and subsequent monitoring~\cite{javierFGCS}.

Secondly, the prevalence of duplicate websites, such as mirrors or phishing sites, can lead to an over-representation of specific services in the data set, making it difficult to portray the distribution of web content~\cite{43} accurately. Phishing is typical due to the infeasibility of checking the authenticity of onion domains due to their anonymous nature~\cite{44}. Mirrors differ from phishing sites since the owners of a web service duplicate them to mitigate DDoS attacks~\cite{17}.

Thirdly, the topics and categories extraction of Tor onion services poses another significant challenge. Manual categorization is time-consuming and labor-intensive, and relying on keyword or bag-of-words approaches may not always yield accurate results. Some studies have utilized machine learning models, such as  Sparse Composite Document Vector, Random Forest, Support Vector Machines, Naive Bayes, and LightGBM, to infer the categories of Tor onion services \cite{17}. Others have employed probabilistic models based on word distributions or third-party software that automatically extract topics without human intervention \cite{15}. However, the challenge of accurately and efficiently categorizing Tor onion services persists, especially when there are advanced techniques available today in the area of Natural Language Processing (NLP) based on Hidden Markov Models, Neural Networks and Bi-directional Encoder Representations from Transformers (BERT)~\cite{khurana2023natural}, or the recent paradigm of prompt learning~\cite{10.1145/3560815}.

This paper proposes a new scalable, efficient, and cost-effective architecture for automated collection and analysis of large-scale contents of the Tor network in near-real time. \addtxt{The purpose is to design and prototype a state-of-the-art solution for large-scale categorization of dark web sites, particularly addressing the three aforementioned challenges identified in the literature}. In particular, the contributions are as follows:

\begin{enumerate}
\item Design, implement and deploy a novel Kubernetes solution for the large-scale near real-time collection and analysis of Tor onion services. The solution includes (i) four different types of sources to early identify fresh onion addresses to address high volatility and capture after dying, (ii) the application of the MinHash LSH (Locally Sensitive Hashing) algorithm to deduplicate dark sites efficiently, and (iii) and categorization employing BERTopic unsupervised approach.

\item Representative research results on extensive experiments on Tor v3 landscape, i) collecting, to the best of our knowledge, the second most representative sample of Tor literature up to date, ii) revealing relevant insights of the considerable proportion of duplicates on the Tor network, and iii) developing, as far as we know, the most representative characterization of onion services based on topics up to date. 
\end{enumerate}

\section{Related work}

\subsection{Tools for analyzing Tor onion services}

Many investigations implement crawlers to explore the dark web. However, this section highlights more sophisticated solutions with additional functions for crawling, as depicted in Table~\ref{table:1}:

\begin{itemize}
    \item \textit{Dark Crawler}~\cite{32} is a crawling tool that starts at user-specified websites, retrieving and analyzing HTML content from discovered onion services, while storing statistics, content, and image hashes. It identified 10,163 onion sites. 
    
    \item \textit{Automated Tool for Onion Labeling (ATOL)}~\cite{46} is an infrastructure that crawls pages twice per day from seed sets and web searches, extracts the main themes, and indexes the results on Elasticsearch. The project analyzed 529 unique Tor pages.
    
    \item \textit{Tor-oriented Web Mining Toolkit}~\cite{Celestini2017} is a Java-based application with four components and two external modules that facilitates massive web crawling, indexing, and text mining. It analyzed 5,144 onion services in six weeks.
    
    \item \textit{Analytical Framework for Darkweb scraping and analysis}~\cite{35} proposes a methodology for scraping targeted marketplaces. This tool is not designed for large-scale monitoring, and AppleScript programs are adapted accordingly for the analysis of each target. Notably, it includes Maltego to extend knowledge about vendors found.

    \item \textit{Docker-based Tor crawler}~\cite{24} uses Docker and paralleled collectors, an analyzer, and a cloud manager to download and analyze dark sites. It monitored the status of 2,527 onion services for five months.
    
    \item \textit{MASSDEAL}~\cite{17} is an exploration and analysis tool built with Python and SQLite that retrieves onions, detects duplicates, classifies resources, discovers new services, and replenishes known services. It was deployed over 105 days and collected 7,831 sites. 
    
    \item \textit{Dark Web Threat Intelligence Analysis (DWTIA) Platform} ~\cite{29} is built with data acquisition, indexing, analysis and visualization modules to process large volumes of data collected from the dark and surface web to identify illicit services. It inspected more than 8,000 sites on the dark web. 

    \item \textit{Dark Web crawling system} \cite{39} uses Selenium, a seed URL collector that periodically visits public dark web services to obtain onion seeds, and a sub URL collector that stores the address, content, and screenshots related to each seed URL. It identified 3,000 onion services in one month and a half.
    
    \item \textit{Black Widow crawler}~\cite{37} is an architecture, which searches, identifies, and indexes secret services, black markets, and criminal patterns using a combination of technologies like Scrapy, Docker, Apache Solr, and MongoDB. The crawler obtained 2,066 onion services in one week. 
    
    \item \textit{Darkweb Monitoring Application}~\cite{50} offers a methodology for analyzing word occurrence, category classification, and visualization of resulting plots. This Python-based tool was tested once with a bulk of 3,000 onion sites.
\end{itemize}

\begin{table}[t!]
\centering
\begin{tabular}{|c|c|c|}
\hline
\textbf{Tool} &  \textbf{Main feature} & \textbf{Onions} \\
\hline
\hline
Dark Crawler \cite{32} & Analyzes HTML & \textbf{10,163}\\
\hline
ATOL \cite{46} & Twice daily crawl & 529\\
\hline
Web Mining Toolkit \cite{Celestini2017} & Java-based & 5,144 \\
\hline
AF for Darkweb \cite{35} & Includes Maltego & -- \\
\hline
Docker Tor crawler \cite{24} & Uses Docker & 2,527 \\
\hline
MASSDEAL \cite{17} & Classifies resources & 7,831 \\
\hline
DWTIA \cite{29} & Large data volume & 8,000 \\
\hline
Web crawl system \cite{39} & Uses Selenium & 3,000 \\
\hline
Black Widow \cite{37} & Uses Scrapy, Docker & 2,066 \\
\hline
Monitoring App \cite{50} & Visualizes plots & 3,000 \\
\hline
\end{tabular}
\caption{Comparison of tools for analysis of Tor onion services}
\label{table:1}
\end{table}

This paper proposes a novel Big Data architecture, improving upon the limited scalability and efficacy of existing methods, for near real-time identification and categorization of Tor sites.

\subsection{Analysis of Tor onion services}

\subsubsection{Temporality and lifetime}

High availability, fixed location, and pseudo-infinite longevity are not expected in Tor. An experiment launched in 2018 determined that approximately 30\% are never reachable, 50\% of the observed onion services were deactivated after 24 hours, and 60\% die after 300 hours~\cite{11}. Another research revealed that only 36\% of the identified sites were alive for 18 weeks~\cite{12}. 

These aspects limit the study of services on the dark web, as seen in Table~\ref{table:2}. For instance, the number of concurrently online services measured was around 1,450 out of more than 7,000 identified addresses~\cite{9}, 7,257 onion links were active out of 198,050~\cite{13}, 30\% was online at least 90\% of the experiment with 47,439 onions identified~\cite{15}, 7K Tor pages were alive out of more than 250K addresses~\cite{31}, or strategies that returned 124,589 addresses with only 3,536 active \cite{26}. 

\change{Similarly, other}{Other} studies highlighted the difficulty in reaching onion services due to their transient nature; only a small percentage were accessible. For example, only 35\% \addtxt{of} sites could be accessed successfully out of 12,882 onion addresses~\cite{22}, 6,227 were online and accessible at the time of the crawl out of 25,742 onion services discovered~\cite{23}, 4,089 were up and responding \change{the}{to} crawler's request out of 15,503~\cite{41}, or 2,527 were open from a total of 25,261 onion addresses~\cite{24}.

\begin{table}[t]
\centering
\begin{tabular}{|c|c|c|c|}
\hline
\textbf{Study} & \textbf{Total onions} & \textbf{Active onions} & \textbf{Portion}\\
\hline
\hline
\cite{9} & 7,000 & 1,450 & 20.7\%\\
\hline
\cite{13} & 198,050 & 7,257 & 3.7\%\\
\hline
\cite{15} & 47,439 & \textbf{14,232} & 30\%\\
\hline
\cite{31} & 250,000 & 7,000 & 2.8\%\\
\hline
\cite{26} & 124,589 & 3,536 & 2.8\%\\
\hline
\cite{22} & 12,882 & 4,509 & \textbf{35\%}\\
\hline
\cite{23} & 25,742 & 6,227 & 24.2\%\\
\hline
\cite{41} & 15,503 & 4,089 & 26.4\%\\
\hline
\cite{24} & 25,261 & 2,527 & 10\%\\
\hline
\end{tabular}
\caption{Tor accessibility in previous studies}
\label{table:2}
\end{table}

In addressing coverage and access issues, the proposed architecture is explicitly configured with different types of live data sources which are continuously monitored.  

\subsubsection{Duplicates, mirrors and phishing sites}

Some studies measure dark content redundancy, as shown in Table~\ref{tab:3}, demonstrating that the same service may be mirrored in different onion addresses:

\begin{itemize}
    \item In an analysis of 2,527 onion services~\cite{24}, the authors decided to consider two onion services to be the same when the cosine similarity of titles and HTML files were over 90\% and 80\%, respectively, reducing the sample to 2,014 unique sites.

    \item A four-month experiment~\cite{17} used the hashes of HTML documents, screenshots and Jaccard distance of titles (over 0.8) to show that approximately 80\% of 7,831 monitored onion services were unique and 20\% had more than one mirror URL (on average, 4.89 onion addresses).

    \item The \change{latter}{Jaccard} distance method was also employed in a large-scale study~\cite{15} of 45,135 dark sites to cluster HTML pages and extract 33,217 duplicates (73.59\%) divided into 1,021 clusters with different average similarity coefficients.
\end{itemize}

The mirroring is specifically addressed in the literature from a cybersecurity perspective: 

\begin{itemize}
    \item Barr-Smith and Wright~\cite{14} studied the imitation of phishing sites using the \textit{html-similarity}\footnote{https://github.com/matiskay/html-similarity} library to estimate the page structure-based similarity. Of 11,533 services, 33.573\% were unique, 45.192\% were imitations (duplicates or clones), and 21.23\% were default pages.

    \item Brenner et al.~\cite{10.1145/3465481.3470026} evidenced that darknet websites can be effectively compared for similarity using category structure features such as HTML-Tag, HTML-Class, and HTML-DOM-Tree, along with metadata features like File Content and Links-To. They demonstrated that out of 258 single vendor shops, 20\% were found to be duplicates, while 31\% exhibited high levels of similarity.

    \item A method to identify phishing candidates~\cite{44} was employed with clustering to a large number of domains based on shared titles and content, analyzing the equivalence of content among onion domains. The study revealed that out of 28,928 onion domains, only 5,718 website groups with distinct content were present, and 901 phishing domains with duplicated content were identified.

    \item Alternatively, Steinebach et al.~\cite{20} demonstrated that comparing images was the best duplication detection for 4,210 onion services in contrast to comparing identical texts or onion addresses.
\end{itemize}

\begin{table}[t]
\centering
\begin{tabular}{|c|c|c|c|}
\hline
\textbf{Study} & \textbf{Technique} & \textbf{Onions} & \textbf{Duplicates} \\ \hline\hline
\cite{24} & Content Similarity & 2,527 & 80\% \\ \hline
\cite{17} & Content Similarity & 7,831 & 20\% \\ \hline
\cite{15} & Content Similarity & 45,135 & 73.59\% \\ \hline
\cite{14} & Structural Similarity & 11,533 & 45.92\% \\ \hline
\cite{10.1145/3465481.3470026} & Structural Similarity & 258 & 20\% \\ \hline
\cite{44} & Content Clustering & 28,928 & 80.23\% \\ \hline
\cite{20} & Image Similarity & 4,210 & \textbf{95.44}\% \\ \hline
\end{tabular}
\caption{Studies deduplicating onion services}
\label{tab:3}
\end{table}

In our framework, we integrate content similarity \addtxt{based on Jaccard score} with MinHash LSH algorithm in the processing pipeline to identify duplicates efficiently and reduce the computing overload of the workflow.

\subsubsection{Topic modeling and document classification}

\begin{table*}[t]
\centering
\newcolumntype{P}[1]{>{\centering\arraybackslash}p{#1}}
\begin{tabular}{|P{1.2cm}|P{5.5cm}|P{1cm}|P{1cm}|p{7.2cm}|}
\hline
\textbf{Study} & \textbf{Type} & \textbf{Onions} & \textbf{Topics} & \textbf{Sample} \\
\hline
\hline
\cite{6} & Manual & 1,171 & 23 & \textit{Child abuse, personal, hacking, blackmarket, porn}\\
\hline
\cite{2} & Manual & -- & 22 & \textit{Drugs, market, fraud, bitcoin, mail, wikis, media}\\
\hline
\cite{23} & Manual & 4,102 & 31 & \textit{Adult, bitcoin, directory, drugs, electronics}\\
\hline
\cite{31,26} & Manual & -- & 25 & \textit{Drugs, credit cards, porn, hacking, cryptos}\\
\hline
\cite{39} & Manual & 3,000 & 15 & \textit{Counterfeits, drugs, hacking, forgery}\\
\hline
\cite{41} & Manual & 2,419 & 14 & \textit{Finance, search engine, drugs, credentials}\\
\hline
\cite{22} & Keywords & \textbf{6,227} & 8 & \textit{Marketplaces, drugs, fraud, cybercrime, porn}\\
\hline
\cite{51} & Keywords & 4,000 & 4 & \textit{Dark markets, socially unjust content, bitcoins}\\
\hline
\cite{13} & BOW + Clustering & 5,883 & 7 & \textit{Directories, default messages, market, bitcoins}\\
\hline
\cite{45} & BOW + Naive Bayes & 2,542 & 6 & \textit{Hacking, drug, develop, porn, news, casino}\\
\hline
\cite{46} & TF-ICF + Keywords + Clustering & 481 & 3 & \textit{Weapons, drugs, hacker}\\
\hline
\cite{17} & TF-IDF-ICS + Keywords + Clustering & 445 & 6 & \textit{Listing, login, market, security, porn, other}\\
\hline
\cite{9} & LDA & 1,481 & 250 & \textit{Trading, financial, politics, intelligence, porn}\\
\hline
\cite{40} & LDA & 3,288 & 9 & \textit{Directory, bitcoin, news, email, multimedia}\\
\hline
\cite{47} & FastText + SCDV + LigthGBM & -- & 15 & \textit{Adult, communication, cryptos}\\
\hline
\cite{50} & TF-IDF + SVM & 3,115 & 4 & \textit{Drug, fake id, hacking, weapon}\\
\hline
\cite{3} & \textit{Mallet} + \textit{uClassify} & 1,813 & 18 & \textit{Drugs, adult, counterfeit, weapons, politics}\\
\hline
\cite{Celestini2017} & \textit{Cogito} software & -- & 17 & \textit{Cybersecurity, fraud, drugs, inf. systems, media}\\
\hline
\end{tabular}
\caption{Studies categorizing Tor onion services}
\label{table:categorization}
\end{table*}

Understanding and categorizing Tor services is crucial due to the vast array of complex and often illicit content within the dark web. Primarily text-based, these services pose substantial challenges in terms of volume and ethical implications for manual interpretation.

Despite the need for automatic approaches to process big amounts of documents and avoid controversial human analysis, early studies utilized manual categorization, as seen in Table~\ref{table:categorization}:

\begin{itemize}
    \item Guitton~\cite{6} examined over a thousand onion services and classified them into 23 categories. 

    \item \rmvtxt{Similarly, }Owen and Savage~\cite{2} conducted a detailed analysis of onion sites and grouped them into 22 classes.

    \item An even larger manual inspection~\cite{23} led to the classification of more than four thousand onion services into 31 categories.

    \item Two studies~\cite{31,26} focused on creating and extending the DUTA dataset, manually tagging onions in 25 classes.

    \item Lee et al.~\cite{39} classified three thousand dark sites into 15 categories\change{, while}{.}
    
    \item \change{another}{Another} project~\cite{41} implemented a Java application to assist in manually categorizing 14 topics.
\end{itemize}

\change{Methodologies}{Some methodologies are} based on the presence of determined keywords in the text or employing the distribution in bag-of-words (BOW):

\begin{itemize}
    \item Kinder et al.~\cite{22} identified eight categories from over six thousand onion services.
    \item Alaidi et al.~\cite{50} followed a similar approach, identifying six subjects.
\end{itemize}

Other \change{methodologies}{procedures} involved clustering based on the frequency of word occurrences\change{, such as the study of}{:}

\begin{itemize}
    \item Sánchez-Rola et al.~\cite{13}\rmvtxt{, which} led to the manual tagging of onion services into seven categories, and an approach that employed Naive Bayes to classify onion sites into six categories~\cite{45}.

    \item The ATOL project~\cite{46} identified three categories based on Term Frequency - Inverse Corpus Frequency (TF-ICF) with clustering.

    \item A similar approach but with Term Frequency — Inverse Document Frequency — Inverse Category Score (TF-IDF-ICS) led to the identification of six classes~\cite{17}. 
\end{itemize}  

Probabilistic and machine learning models have been less employed in Tor research for topic extraction:

\begin{itemize}
    \item Latent Dirichlet Allocation (LDA) was used in Tor-based projects like one that identified 250 themes~\cite{9}, and another that identified nine categories~\cite{40}.

    \item Kawaguchi and Ozawa~\cite{47} used a combination of FastText, Sparse Composite Document Vector (SCDV), and LightGBM to classify onion services into 15 categories.

    \item Nair and Kannimoola~\cite{51} used automatic labeling and classification methods to discover four categories in dark sites.
\end{itemize}

Finally, applied projects use third-party software for automated extraction of topics:

\begin{itemize}
    \item Biryukov et al.~\cite{3} used Mallet and uClassify tools to derive 18 categories.

    \item  Celestini and Guarino~\cite{Celestini2017} used the Cogito semantic engine to extract tags based on their topic taxonomy, resulting in 17 classes.
\end{itemize}

However, the current state of the art in topic modeling and document classification is marked by significant advancements in neural networks and word embeddings~\cite{ijcai2021p638}, techniques not present in the categorization of Tor services reviewed so far. Traditional methods such as LDA have been enriched with word embeddings to enhance their effectiveness~\cite{8029739}.

Additionally, models primarily built around embeddings have gained prominence, showcasing the potential of such methodologies for topic modeling~\cite{dieng-etal-2020-topic}. Another emerging trend is simplifying the topic-building process by clustering word and document embeddings~\cite{DBLP:journals/corr/abs-2008-09470}. One of the most recent developments is BERTopic~\cite{grootendorst2022bertopic}, a clustering approach that integrates a class-based variant of Term Frequency-Inverse Document Frequency (TF-IDF) to create topic representations that are being increasingly adopted~\cite{Hanley_Kumar_Durumeric_2023,egger2022topic}.

In order to drive a significant shift in dark site categorization, our architecture integrates the aforementioned BERTopic technique. 

\section{Architecture for identification and categorization of Tor sites}

\begin{figure*}[t!]
    \centering
    \includegraphics[scale=0.55]{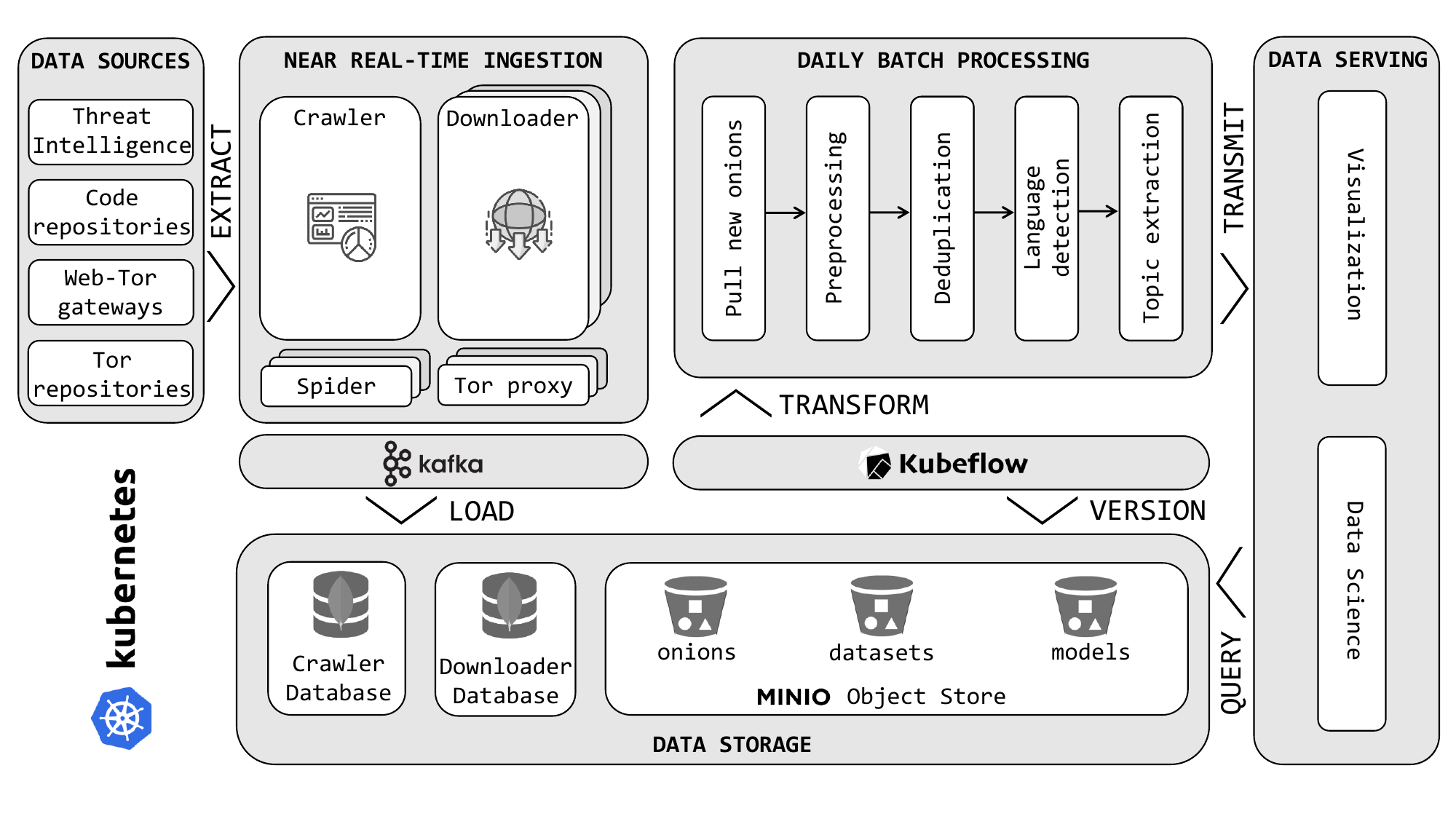}
    \caption{Architecture for the continuous collection and analysis of Tor onion services}
    \label{fig:architecture}
\end{figure*}

The proposed system for the automatic collection and analysis of onion services is divided into separate layers, components, and technologies, illustrated in Figure~\ref{fig:architecture}. Adopting a microservices architecture managed via Kubernetes is fundamental for addressing critical aspects such as scalability, resilience, and maintainability. This decision was motivated by the flexibility and scalability provided by the microservices approach, which allows individual components to be independently scaled according to demand. As a proven orchestration platform, Kubernetes ensures reliable deployment, networking, and scaling of these microservices, improving the overall resilience. Additionally, the loose coupling inherent in the microservices design simplifies maintenance, as each service can be independently updated or replaced without impacting the entire system.

The architecture includes a real-time ingestion layer with a pool of spiders to crawl multiple data sources and extract undiscovered onion addresses. A group of downloaders saves the HTML file of each Tor site using Tor proxies. At the end of the day, a batch processing job analyzes the accumulated onion services, identifies duplicates, classifies languages, and extracts topics. The architecture follows the ELT (Extract, Load, Transform) paradigm. The code is publicly available at \href{https://github.com/javier-pg/dark-web-architecture}{\textsf{github.com/javier-pg/dark-web-architecture}} and detailed information on each layer and component is provided in the following sections.

\subsection{Data Sources}

The role of data sources is pivotal to facilitating the constant acquisition of new onion addresses (complex links composed of 56 base32-coded characters with the “.onion” top-level domain). In this regard, four types of data sources are considered to ensure a frequent supply of new Tor sites to be ingested and analyzed:

\begin{itemize}

\item \textit{Threat intelligence}. Provides valuable information on known threat actors, their tactics, techniques, and procedures (TTPs), and the infrastructure they use. To the best of our knowledge, we are the first to inspect indicators of compromise (IoC), such as domain names, hostnames, URLs, DNS records and advertising links to match onion domains. The latter is used by cybercriminals to host anonymous illicit services or command and control servers on the Tor network. The architecture visits six different threat intelligence feeds\footnote{Including AlienVault OTX, MalwareWorld, Maltrail, Notracking blocklists, Little Snitch blocklists, and USOM (Computer Emergency Response Team of Turkey) blocklists.}.

\item \textit{Code repositories}. Versioning platforms offer a rich resource of open-source software projects, scripts, files, and configurations. For the first time among related works, as far as we know, GitHub repositories are inspected to extract onion addresses hard-coded by developers.

\item \textit{Web-Tor gateways}. Services that act as proxies between traditional explorers and the Tor network, such as Tor2web, dump Tor pages to the surface web. This fact can be exploited with the keyword search algorithm~\cite{16}, based on the dorking \textit{site} operation, such as \textit{site:tor2web.org}) to extract onion domains indexed in search engines under Web-Tor proxies. The architecture considers 19 different alternatives of proxy services\footnote{Including onion.city, onion.direct, onion.gq, onion.link, onion.nu, onion.pw, onion.top, onion.pet, tor2web.to, tor2web.fyi, tor2web.io, onion.ws, onion.foundation, onion.dog, onion.moe, onion.re, onion.ly, onion.pet, and onion.cab.}.

\item \textit{Tor repositories}. Compilations of Tor links are widely used in the literature to gather onion addresses due to the accuracy, speed, and simplicity of consulting ready-to-use compilations~\cite{javierFGCS}. The architecture is configured with 13 different repositories\footnote{Including Ahmia, OnionRanks, TheHiddenWiki, Dark.Fail, DarknetLive, TheDeepSearches, FreshOnions, Torch, Tor Links, Tor66, OnionLand Search, TorDex, and H-Indexer.}.
\end{itemize}
 
The four types of onion sources provide a quantity and variety of onions daily, reducing the search bias potential of this type of platform.

\subsection{Near Real-time Ingestion}

The Near Real-time Ingestion layer extracts onion addresses from data sources and downloads the HTML document. The onion service is not immediately identified as soon as it appears online. Instead, it is immediately detected after it is advertised in the data sources and accessed by the system.

\subsubsection{Crawler and spiders}

The architecture deploys a virtual instance of Crawlab\footnote{https://github.com/crawlab-team/crawlab}, a Web Crawler Management Platform that runs custom web crawlers. This platform provides a user-friendly interface for importing, running, managing, and monitoring spiders, making them traceable, scalable, and stable. Our solution imports, configures and schedules four Python spiders:

\begin{itemize}
    \item \textit{Threat intelligence spider}: Scrapes the six feeds based on regular expressions of the listings every six hours.
    
    \item \textit{Code repositories spider}: Searches with Grep.app\footnote{https://grep.app} and Sourcegraph\footnote{https://about.sourcegraph.com/code-search} for GitHub content matching the string \textit{d.onion} (the last compulsory characters of Tor domains), every six hours.
    
    \item \textit{Web-Tor gateways spider}: Uses DuckDuckGo to identify indexed onion links from the 19 web-Tor proxies every six hours. Programmatic search permissions were requested from Google and Bing without success.
    
    \item \textit{Tor repositories spider}: Employs Scrapy\footnote{https://scrapy.org/} to crawl the 13 repositories with a depth of 3 and a download delay of 0.2 seconds every twelve hours.
\end{itemize}

Each spider is independent, can be easily extended to request \addtxt{for} more resources, and maintains the information of its operations in the Crawler Database. The execution frequencies are configurable, and current decisions are based on tests for a good trade-off between intensive search and efficacy to capture new inclusions at an early stage, especially for those volatile onion addresses that tend to disappear or be eliminated quickly in very short periods.

\subsubsection{Downloaders and Tor proxies}

The downloader is a Python-based Kafka consumer that retrieves the new onion addresses, connects to the Tor network through the SOCKS proxy, and saves the associated raw HTML page (without media and Javascript) and the MinIO object storage instance. Particularly, the architecture runs a set of downloaders scaled with Kubernetes to the desired number of parallel replicas (ten in our experimental deployment). Similarly, it scales Tor proxies according to the needs (five in our running cluster).

A Kafka source connector on the Crawler Database creates one Kafka topic per data source to forward new Tor links to the downloaders automatically. Therefore, the latter receives the new onion addresses when the Crawler inserts new entries in each data source collection (event-driven change data capture). Each replica receives one Kafka partition per topic.

On the one hand, MinIO saves the raw HTML file in the \textit{onions} bucket. On the other hand, the Downloader Database keeps track of every onion in the system that has been found. In this sense, an HTML page will only be downloaded if it has not been saved before.

\subsection{Daily Batch Processing}

The pipeline of this layer is automated and virtualized using Kubeflow Pipelines SDK v2 to dynamically deploy and execute batch processing at the end of each day, enabling transparent coordination and data sharing between the different phases. 

\change{This approach means that}{With this approach,} resources are not reserved throughout the day but are automatically demanded only for a limited period to build up the pipeline, analyze the new onions, and destroy the associated containers when finished. The following are the steps involved in the daily batch processing:

\begin{figure*}[t]
    \centering
    \includegraphics[scale=0.7]{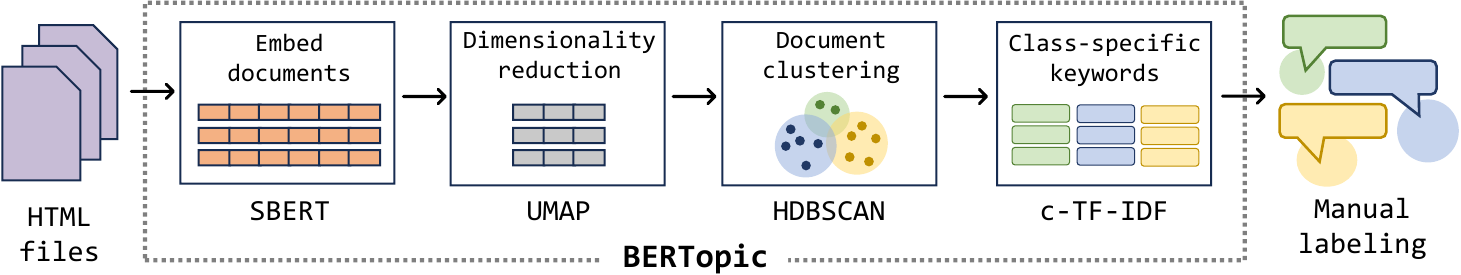}
    \caption{BERTopic methodology to classify onion preprocessed HTML files}
    \label{fig:BERTopic}
\end{figure*}

\subsubsection{Pull new onion sites}
The first action is to retrieve HTML documents from the \textit{onions} bucket of the MinIO store by date. The HTML is parsed with BeautifulSoup, and the valuable part is extracted with Trafilatura~\cite{barbaresi-2021-trafilatura}, the framework that gave us the most refined results.

Automatically extracting meaningful text from raw HTML content and discarding tags, metadata, descriptions, footers, or style code, is critical particularly in the context of the dark web. Unlike the standard surface pages that adhere to well-established best practices, dark sites exhibit a less sophisticated implementation, limiting parsing frameworks' performance. It is worth mentioning that some dark pages are bad-encoded or empty.

\subsubsection{Preprocessing of textual content}
The meaningful text is preprocessed using the NLTK sentence tokenizer to remove worthless text, discarding onion addresses, PGP keys, email addresses, monetary values, URLs, bitcoin addresses, and HTML tags. Line breaks, tabulations, special characters, contiguous duplicated characters, words, and sentences are \addtxt{also} removed. 

By using document embeddings, there is no need to do more NLP preprocessing to understand the general topic of the document. The result is a cleaned and compacted text.

\subsubsection{Efficient deduplication of onion sites}
The preprocessed pages are deduplicated based on document similarity in the absolute dataset. \change{Exact duplicates are identified through the exact matching of texts, and near-duplicates are grouped using MinHash LSH (Jaccard threshold of 0.9 and 128 permutation functions). The latter  is quick, efficient and scalable. It uses hashing to map high-dimensional data points to buckets and identifies potential matches by their MinHash signatures, reducing the number of total similarity comparisons to perform. The longest document in each bucket represents that group.}{Firstly, exact duplicates are identified through the identical matching of texts. For near deduplication of the rest of the documents, MinHash LSH~\cite{gionis1999similarity} is adopted to map the set into smaller buckets to reduce the number of comparisons to perform. In particular, Hash signatures are calculated with MinHash algorithm for each document, being grouped via LSH. Finally, each bucket has a subset to compare, the content similarity of documents is performed with a similarity measurement}.

\addtxt{After testing two of the most common measurements, Cosine Similarity and Jaccard Similarity \cite{gomaa2013survey}, the selection of the latter with a threshold of 0.9 and 128 permutation functions emerged as the configuration for the best performance.}

\subsubsection{Language detection}
The never-seen documents are analyzed to extract the language of the new onions using the language model\footnote{lid.176.bin} from Facebook AI's \textit{fastext} framework, \addtxt{one of the best alternatives in terms of performance and scalability}~\cite{joulin2016bag}.

\subsubsection{Topic extraction}
\rmvtxt{The English onion services are categorized using BERTopic modeling~[45], as shown in Figure 2. Particularly, SBERT~[53] multilingual model\footnote{paraphrase-multilingual-MiniLM-L12-v2} is used to embed documents to a high-dimensional subspace, UMAP~[54] for reducing the dimensionality, HDBSCAN~[55] for document clustering, and c-TF-IDF~[56] for topic keywords. The resulting clusters are manually labeled with categories by the authors. Secondly, the model is deployed to predict new unseen onion services.}

\addtxt{To categorize English onion services, we employ BERTopic modeling~\cite{grootendorst2022bertopic}, a technique that provides insights into the underlying themes present in a collection of documents. Figure \ref{fig:BERTopic} illustrates the process.}

\addtxt{The first step involves leveraging the SBERT~\cite{reimers-gurevych-2019-sentence} multilingual model, specifically the \textit{paraphrase-multilingual-MiniLM-L12-v2} variant. This model transforms documents into a high-dimensional subspace, capturing semantic relationships and contextual information. Subsequently, UMAP~\cite{becht2019dimensionality} is utilized to reduce the dimensionality of the embedded documents. This step is crucial for visualizing and understanding the data in a more manageable space.}

\addtxt{For document clustering, we employ HDBSCAN~\cite{mcinnes2017hdbscan}, a hierarchical density-based clustering algorithm. This algorithm not only identifies clusters but also allows for the detection of outliers, enhancing the robustness of the categorization process. In order to extract meaningful topic keywords within each cluster, we turn to c-TF-IDF~\cite{ozgur2005text}. This technique emphasizes terms that are not only frequent within a specific document but also discriminative across the entire corpus, providing a more nuanced representation of topics.}

\addtxt{The resulting clusters are manually labeled with categories by the authors, providing a human-in-the-loop validation of the algorithm's outputs. This step adds a layer of interpretability and ensures that the generated categories align with real-world semantic distinctions.}

\vspace{0.5cm}

In the Kubeflow pipeline, each \change{component}{phase discussed above} is deployed on a Kubernetes Pod to carry out the programmed task and pass the output to the next component, but also versions the newly processed data on MinIO \textit{datasets} bucket. Therefore, each component i) generates the daily dataset, statistical plots, and metrics, and ii) merges the new dataset with the\rmvtxt{ general one} updated general dataset with global statistical plots and metrics\footnote{Datasets are modeled with Kubeflow Output Artifacts, plots with Kubeflow Output HTML, metrics with Kubeflow Output Metrics}. The Pods are automatically removed when associated activities have finished.

\subsection{Data Storage}

This layer is responsible for persistently storing the data that results from the architecture. The system uses two non-relational MongoDB databases for ingestion and MinIO, a high-performance Kubernetes-native object storage, for processing.

\subsubsection{Crawler Database}
This database stores the data related to crawling functions, configurations, and results. Specifically, each of the four source-based spiders saves its findings in a separate collection, including the \textit{onion address} (a unique index), \textit{advertiser}, and \textit{identification timestamp}. 

The four collections are configured with the MongoDB Kafka Source Connector to raise events with new insertions and feed the downloaders.

\subsubsection{Downloader Database}
This database maintains the details to manage the download of the HTML files such as the \textit{onion address}, \textit{downloaded} (whether the onion has been downloaded), \textit{downloaded timestamp} (date when the onion was downloaded), and four booleans for each data source indicating where the onion has been identified.

\subsubsection{MinIO Object Store}
This component is used to store various raw artifacts accessed throughout the architecture. The \textit{onions} bucket contains every downloaded HTML page, and the \textit{datasets} bucket keeps the incremental versions of the preprocessed, deduplicated, languages, and topic datasets generated by the daily pipeline. Both buckets index objects by date (\textit{yyyy/mm/dd}) for easy access by batch processing. \change{The language and topic models are stored in the \textit{models} bucket.}{On the other hand, the \textit{models} bucket stores the language and topic models to be used.}

Notably, all this persistence through copies in MinIO allows replication of the pipeline from any point in time in the occurrence of a catastrophic event.

\subsection{Data Serving}

The system provides a comprehensive suite of metrics and visualizations covering daily and global operations, all thanks to the user-friendly Kubeflow visual interface. Additionally, the architecture exposes the MinIO instance that enables data scientists to perform potential complex operations easily.

\section{Experimental results}

The architecture is implemented on a physical server with four virtual machines, all integrated with a self-hosted Kubernetes cluster (one master node and three worker nodes). The storage is provisioned with NFS to ensure reliable data storage and retrieval.

In addition to the core components, the system incorporates management with Kubernetes Dashboard, continuous integration and deployment (CI/CD) with GitLab Agent and Runners, and node resource monitoring with Prometheus and Grafana. In total, the system runs 66 pods to ensure efficient and seamless operation.

\begin{figure*}[t!]
    \centering
    \includegraphics[scale=0.93]{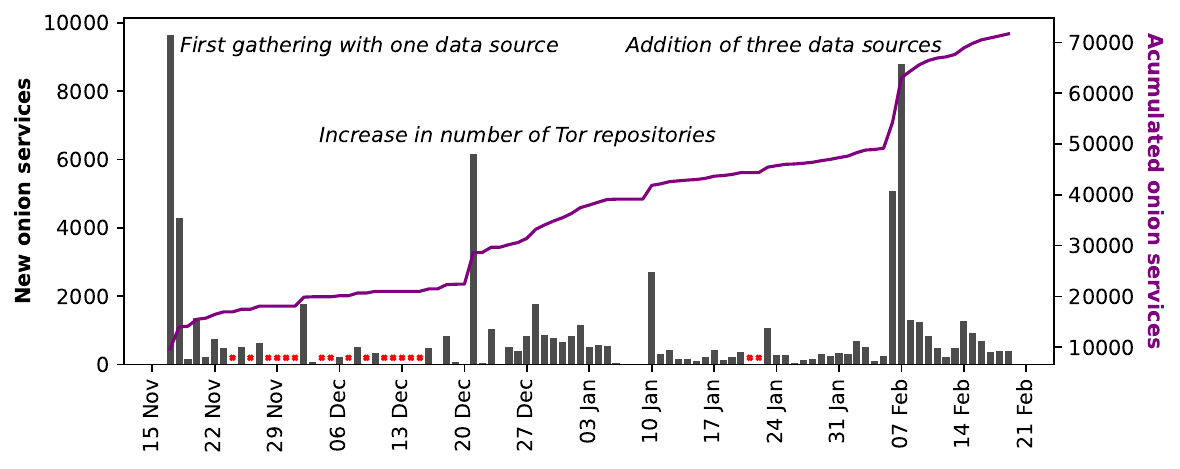}
    \caption{Identification and analysis of new onion services}
    \label{fig:collection}
\end{figure*}

\subsection{Ingestion and preprocessing of onion services}

\subsubsection{Ingestion}
Our solution operated from November 17, 2022, to February 19, 2023\footnote{Except for January 21 and 22, due to technical problems in the deployment}, and uncovered a total of 72,045 active Tor onion sites, as shown in Figure~\ref{fig:collection}. In the first month of execution, we used only one resource as a data source (Ahmia), which resulted in days without updates and an average rate of 659.4 onions/day. On December 21, we added the rest of the listings of the Tor repository source, which increased the rate to 816.7 onions/day. On February 6 and 7, we included the other three types of data sources (threat intelligence, code repositories, and Web-Tor gateways), which doubled the gathering to 1608.9 onions/day.

Table~\ref{tab:sources} shows the number of identified sites per data source, with Tor repositories being the most successful, with 71,742 active onions discovered (771.4 per day). Interestingly, 303 onions are identified in alternative data sources, with 29 uniquely discovered by threat intelligence, 153 in code repositories, and 65 through Web-Tor gateways.

\subsubsection{Preprocessing}
The raw dataset contains 72,045 HTML documents. The platform has identified 153 bad-encoded documents and 149 empty onions along the daily batches. The preprocessing step causes 39 new HTML files to be empty as well.

The final preprocessed sample consists of 71,704 documents, which are exactly deduplicated with perfect hash matching and near deduplicated with MinHash LSH. In particular, there are 56,674 exact duplicates (78.7\%) and 10,640 near duplicates (14.8\%)\footnote{Related to the efficiency of MinHash LSH, the near deduplication of onions processed last day considering 70K processed documents took around 5 seconds.}, reaching a total sum of 67,314 duplicates (93.5\%). The deduplicated dataset contains 4,390 unique onion services, a 6.1\% of the dataset. 

\begin{table}[t!]
\centering
\begin{tabular}{c c c c}
\hline
\textbf{Source} & \textbf{Days} & \textbf{Total onions} & \textbf{Active onions}\\
\hline
Threat Int. & 13 & 952 & 296 (31.1\%)\\

Code repos. & 13 &  1,741 & 819 (47\%)\\

Gateways & 13 & 759 & 677 (89.2\%)\\

Tor repos. & 93 & 78,656 & 71,742 (91.2\%)\\ 
\textbf{Total} &  93 & 80,049 & \textbf{72,045} (90\%)\\
\hline
\hline
\textbf{Bad/Empty} & \textbf{Exact dup.} & \textbf{Near dup.} & \textbf{Unique} \\
\hline
341  & 56,674  & 10,640  & 4,390  \\
(0.4\%) & (78.7\%) & (14.8\%)& (6.1\%) \\
\hline
\hline
\end{tabular}
\caption{Ingestion per data source and preprocessing stats}
\label{tab:sources}
\end{table}

\subsection{Extraction of languages and topics}

\subsubsection{Language distribution}
The pipeline is configured to extract the predominant language of each unique document. \addtxt{More than 30 different languages were detected, and the five most predominant among the 4,390 sites were the following:}

\begin{itemize}
    \item \addtxt{English: 3,869 (88.1\%)}
    \item \addtxt{Russian: 126 (2.9\%)}
    \item \addtxt{French: 84 (1.9\%)}
    \item \addtxt{German: 61 (1.4\%)}
    \item \addtxt{Spanish: 50 (1.1\%)}
\end{itemize}

\rmvtxt{Of the 4,390 sites, 3,869 (88.1\%) are in English, 126 (2.9\%) are in Russian, 84 (1.9\%) are in French, 61 (1.4\%) are in German, and 50 (1.1\%) are in Spanish, among more than 30 languages.}

The distribution changes with the propagation of language to the \addtxt{dataset of 71,704 documents with} duplicates: 

\begin{itemize}
    \item \addtxt{English: 69,499 (96.9\%)}
    \item \addtxt{German: 1,060 (1.5\%)}
    \item \addtxt{Romanian: 354 (0.5\%)}
    \item \addtxt{Russian: 220 (0.3\%)}
    \item \addtxt{Spanish: 157 (0.2\%)}
\end{itemize}

\rmvtxt{of the 71,704 documents, 69,499 (96.9\%) are in English, 1,060 (1.5\%) are in German, 354 (0.5\%) are in Romanian, 220 (0.3\%) are in Russian, and 157 (0.2\%) are in Spanish, among others.} 

Given the representativeness of the English language and the fact that BERTopic works best with it~\cite{grootendorst2022bertopic}, the architecture applies the topic model only to English documents\footnote{A multilingual sentence-transformer is used due to alternative languages present in English documents.}.

\subsubsection{Low-level topics}

The BERTopic process groups the 3,869 English unique onions in 35 different clusters, as shown in Table~\ref{tab:ltopics}, together with the words extracted from the documents under class-based TF-IDF (c-TF-IDF) and the label manually placed. The set of keywords may not be representative enough of each grouping, so the authors do the labeling manually based on the keywords, documents, and embedding distances.

As a result, the top-5 most common are sexual content (1,086), repositories and search engines (877), carding (394), cryptocurrencies (369), and hacking (169). In general, this indicates the predominance of apparently illicit buying and selling sites.

\begin{table*}[t!]
\centering
\begin{tabular}{c c c l l}
\hline
\multicolumn{1}{c}{\textbf{Cluster}} & \multicolumn{1}{c}{\textbf{Onions}} & \multicolumn{1}{c}{\textbf{Ratio}} & \multicolumn{1}{c}{\textbf{Top-5 class-specific words}} & \multicolumn{1}{c}{\textbf{Manual low-level topic label}}\\
\hline
\hline
0 & 1,086 & 28.1\% & \texttt{porn, video, sex, teen, girl} & Sexual content\\

1 & 877 & 22.7\%  & \texttt{csv, uniquevoteshashes, distribution, git, diff} & Repositories and search engines\\

2 & 394 & 10.2\%  & \texttt{card, covid, transfer, money, paypal} & Carding\\

3 & 369 & 9.5\%  & \texttt{bitcoin, ethereum, monero, crypto, tether} & Cryptocurrencies\\

4 & 169 & 4.4\%  &  \texttt{hacking, whatsapp, phone, hire, hackers} & Hacking\\

5 & 136 & 3.5\%  & \texttt{cocaine, drugs, quality, shipping, lsd} & Drug marketplaces\\

6 & 82 & 2.1\%  & \texttt{returns, patch, mcdonald, human, ronald}  & Personal websites and blogs\\

7 & 76 & 2\%  &  \texttt{male, coconut, news, read, oil} & News and media\\

8 & 47 & 1.2\%  &\texttt{guns, rifle, magazine, gun, pistol} & Firearm marketplaces\\

9 & 44 & 1.1\%  & \texttt{download, upload, file, image, mins} & Image and file hosting\\

10 & 43 & 1.1\%  &  \texttt{gz, tar, live, jan, sqxz} & 'Index of' pages\\

11 & 40 & 1\%  & \texttt{nitter, irg, vehicles, rcmp, totoro} & One-line pages\\

12 & 37 & 1\%  &  \texttt{counterfeit, money, banknotes, bills, fake} & Counterfeits\\

13 & 35 & 0.9\%  & \texttt{documents, passport, fake, license, certificate} & Passports and certificates\\

14 & 34 & 0.9\%  & \texttt{login, javascript, password, sign, register} & Javascript pages\\

15 & 32 & 0.8\%  & \texttt{error, nginx, server, forbidden, var} & Error pages\\

16 & 31 & 0.8\%  &  \texttt{dumps, pin, card, track, fullz} & Card dumps and fullz\\

17 & 31 & 0.8\%  &  \texttt{iphone, apple, delivery, phone, samsung} & Mobile and device marketplaces\\

18 & 27 & 0.7\%  &  \texttt{debian, backports, ports, translations, key} & Debian community\\

19 & 26 & 0.7\%  & \texttt{hitman, killer, hire, person, job} & Hitman services\\

20 & 24 & 0.6\%  &  \texttt{escrow, seller, buyer, dispute, transaction} & Escrow services\\

21 & 23 & 0.6\%  &  \texttt{token, finance, coin, protocol, network} & Crypto swapping and exchanges\\

22 & 22 & 0.6\%  & \texttt{bank, logs, transfer, account, money} & Banking\\

23 & 21 & 0.5\%  &  \texttt{privacy, cwtch, security, server, metadata} & Privacy-preserving services\\

24 & 20 & 0.5\%  & \texttt{rog, gaming, performance, strix, zephyrus} & Hardware marketplaces\\

25 & 19 & 0.5\%  & \texttt{dpp, login, admin, password, png} & Login and register pages\\

26 & 18 & 0.5\%  &  \texttt{debian, debconf, conference, info, talks} & Debian conferences\\

27 & 18 & 0.5\%  &  \texttt{incest, eddie, gallery, mindy, tiger} & Violent content\\

28 & 16 & 0.4\%  &  \texttt{cvv, dumps, fullz, bases, bazar} & CVV marketplaces\\

29 & 15 & 0.4\%  &  \texttt{redirected, queue, wait, refresh, automatically} & Redirecting pages\\

30 & 13 & 0.3\%  &  \texttt{ddos, mirror, anti, protection, bank} & DDoS protection pages\\

31 & 11 & 0.3\%  & \texttt{matches, fixed, cup, fifa, betting}  & Betting services\\

32 & 11 & 0.3\%  &  \texttt{instagram, hack, facebook, database, accounts} & Data leaks\\

33 & 11 & 0.3\%  &  \texttt{post, topics, view, neneh, subforums} & Forums\\

34 & 11 & 0.3\%  & \texttt{ammo, min, read, collection, backup} & Generic markets\\
\hline
\textbf{Total} & 3,869 & 100 & \\
\hline
\hline
\end{tabular}
\caption{Low-level topics of unique onion sample}
\label{tab:ltopics}
\end{table*}

\rmvtxt{On the other hand, Figure 4 illustrates the hierarchical structure of topics based on the cosine distance between their embeddings. The closer the cosine distance is to zero, the more similar the topics are.}

\begin{figure*}[ht!]
    \centering
    \includegraphics[scale=0.7]{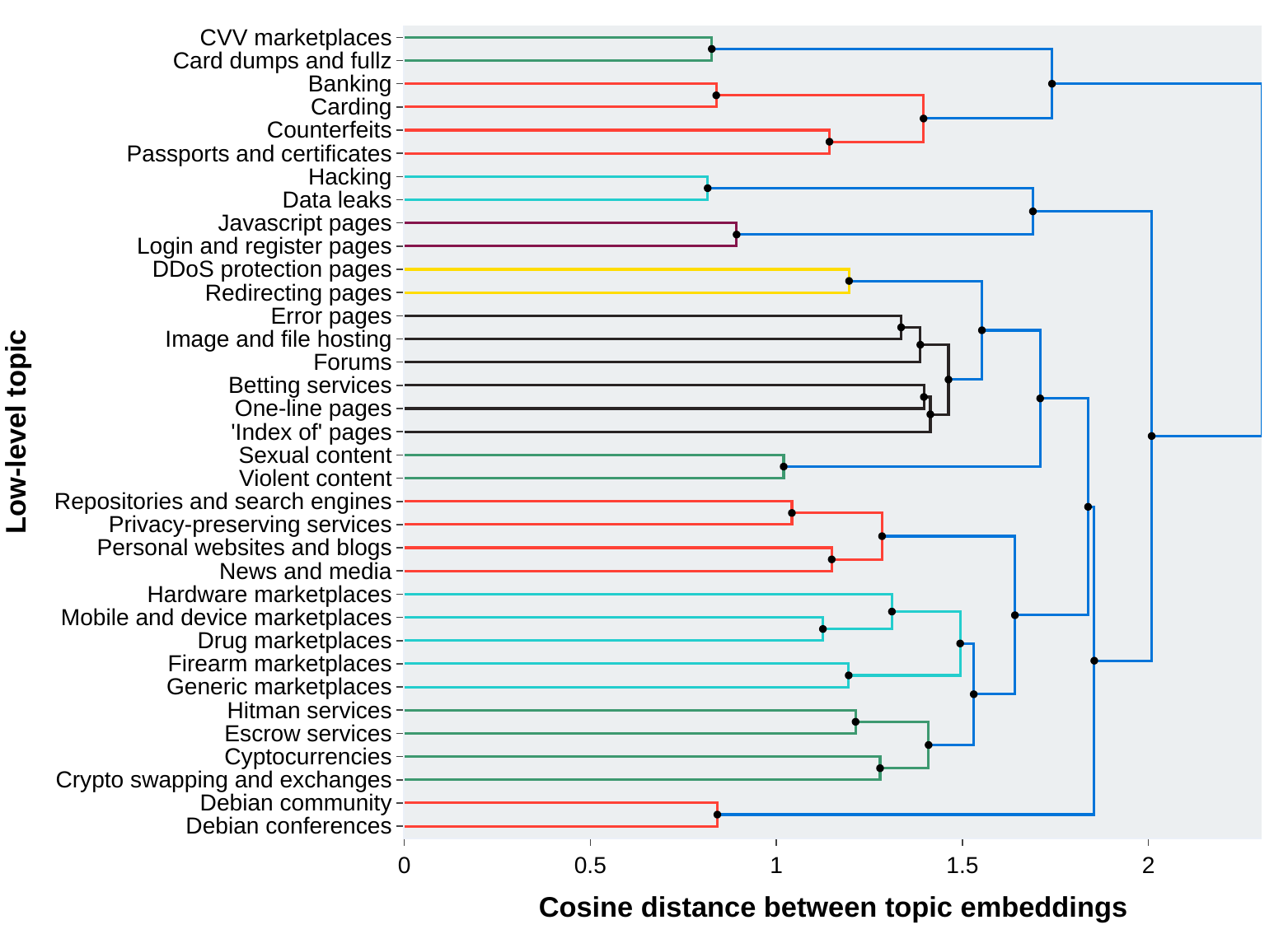}
    \caption{Hierarchical clustering based on the cosine distance matrix between topic embeddings.}
    \label{fig:hierarchical}
\end{figure*}

\addtxt{On the other hand, Figure~\ref{fig:hierarchical} illustrates the hierarchical structure of topics based on the cosine distance between their embeddings. The cosine distance is calculated by measuring the cosine of the angle between two vectors. Specifically, for each pair of topic embeddings, the cosine distance is computed as the dot product of the vectors divided by the product of their magnitudes.}

\addtxt{While alternative distance metrics, such as Euclidean distance and Jaccard similarity, were considered, the cosine distance emerged as the preferred choice for our task. Euclidean distance measures the straight-line distance between two points in space, which may be sensitive to the magnitude of the vectors and less effective in capturing semantic similarity. Jaccard similarity, on the other hand, focuses on the intersection and union of sets, which might not fully capture the nuances of semantic relationships.}

\addtxt{The cosine distance, in contrast, emphasizes the directional similarity between vectors, making it robust to variations in document lengths. This characteristic is particularly advantageous in our context, where documents may differ in terms of length but still convey similar semantic content. Consequently, the choice of cosine distance enhances the reliability of our hierarchical topic structure, providing a more accurate representation of the semantic relationships between different categories.}

In the tree of Figure~\ref{fig:hierarchical}, related topics tend to be grouped and have smaller cosine distances, as indicated by the colors. For example, there is \change{the}{a} closeness between topics related to i) credit cards or documents, ii) hacking and data leaks, iii) sexual and violent content,  \addtxt{or} iv) marketplaces, \addtxt{hiring} services or crypto services\rmvtxt{, among others}.

\subsubsection{High-level topics}

Considering the cosine distances between topic embeddings and author's expertise, the fine-grained clusters are manually merged into fewer groups to facilitate the analysis. As shown in Table~\ref{tab:htopics}, 11 high-level topics \change{raises}{raise} \addtxt{from the 35 low-level categories}:

\begin{enumerate}
    \item \addtxt{Sexual and violent content: \textit{sexual content}, and \textit{violent content}.}
    
    \item \addtxt{Repositories and search engines: \textit{repositories and search engines}.}
    
    \item \addtxt{Carding: \textit{CVV marketplaces}, \textit{card dumps and fullz}, \textit{banking}, and \textit{carding}.}

    \item \addtxt{Cryptocurrencies: \textit{cryptocurrencies}, and \textit{crypto swapping and exchanges}.}

    \item \addtxt{Marketplaces: \textit{hardware}, \textit{mobile and device}, \textit{drug}, \textit{firearm}, and \textit{generic marketplaces}.}

    \item \addtxt{Media, forums and personal websites: \textit{media}, \textit{personal websites and blogs}, \textit{news and media}, \textit{debian community}, and \textit{debian conferences}.}

    \item \addtxt{Navigation pages: \textit{javascript}, \textit{login and register}, \textit{DDoS protection}, \textit{redirecting}, \textit{error}, \textit{one-line}, and \textit{'index of' pages}.}

    \item \addtxt{Hacking: \textit{data leaks}, and \textit{hacking}.}

    \item \addtxt{Counterfeits: \textit{passports and certificates}, and \textit{counterfeits}.}

    \item \addtxt{Privacy-preserving services: \textit{image and file hosting}, and \textit{privacy-preserving services}.}

    \item \addtxt{Hiring services: \textit{betting}, \textit{hitman services}, and \textit{escrow services}.}
\end{enumerate}

\rmvtxt{sexual and violent content (sexual content and violent content); repositories and search engines (repositories and search engines); carding (CVV marketplaces, card dumps and fullz, banking, carding), cryptocurrencies (cryptocurrencies, crypto swapping and exchanges); marketplaces (hardware, mobile and device, drug, firearm and generic marketplaces); media, forums and personal websites (media,personal websites and blogs, news and media, debian community and debian conferences); navigation pages (javascript, login and register, DDoS protection, redirecting, error, one-line and index of pages); hacking (data leaks, hacking); counter-
feits (passports and certificates, counterfeits); privacy preserving services (image and file hosting, privacy preserving services); and hiring services (betting, hitman services and escrow services).}

From this point, we focus on high-level topics, whose distribution both in the unique dataset and total dataset with duplicates is also presented in Table~\ref{tab:htopics}. Considering the 3,869 unique sites without mirrors, the top-5 high-level topics sum 79.6\%: 

\begin{enumerate}
    \item \addtxt{Sexual and violent content: 1,104 (28.5\%)}
    \item \addtxt{Repositories and search engines: 877 (22.7\%)}
    \item \addtxt{Carding: 463 (12\%)}
    \item \addtxt{Cryptocurrencies: 392 (10.1\%)}
    \item \addtxt{Marketplaces: 245 (6.3\%)}
\end{enumerate}

\change{sexual and violent content (1,104), repositories and search engines (877), carding (463), cryptocurrencies (392) and marketplaces (245). In order of prevalence, the}{The} rest of the topics are media, forums and personal websites, navigation pages, hacking, counterfeits, privacy-preserving services and hiring services.

However, if we consider the replication of onion services, the total distribution of topics becomes very unbalanced, increasing even more the presence in the dark web of the first five topics to 95,3\%: 

\begin{enumerate}
    \item \addtxt{Sexual and violent content: 33,530 (48.2\%)}
    \item \addtxt{Repositories and search engines: 19,098 (27.5\%)}
    \item \addtxt{Carding: 6,912 (9.9\%)}
    \item \addtxt{Cryptocurrencies: 4,829 (6.9\%)}
    \item \addtxt{Marketplaces: 1,859 (2.7\%)}
\end{enumerate}

\rmvtxt{sexual and violent content (33,530), repositories and search engines (19,098), carding (6,912), cryptocurrencies (4,829) and marketplaces (1,859).} These types of onion services exhibit a significant mirroring, particularly 29.37, 20.78, 13.93, 11.32 and 6.59 duplicates per single onion, respectively.

\begin{table*}[t!]
\centering
\begin{tabular}{l l c c c}
\hline
\multicolumn{1}{c}{\textbf{High-level topic}} & \multicolumn{1}{c}{\textbf{Low-level topics}} & \multicolumn{1}{c}{\textbf{Uniques}} & \multicolumn{1}{c}{\textbf{Mirrors}} & \multicolumn{1}{c}{\textbf{Total}}\\
\hline
\hline
\multirow{2}{*}{Sexual and violent content} & Sexual content & \multirow{2}{*}{1,104 (28.5\%)} & \multirow{2}{*}{32,426} & \multirow{2}{*}{33,530 (48.2\%)}\\
                                             & Violent content &  & \\

\hline
Repositories and search engines              & Repositories and search engines & 877 (22.7\%) & 18,221 & 19,098 (27.5\%)\\

\hline
\multirow{4}{*}{Carding}                     & CVV marketplaces & \multirow{4}{*}{463 (12\%)} & \multirow{4}{*}{6,449} & \multirow{4}{*}{6,912 (9.9\%)}\\
                                             & Card dumps and fullz &  & \\
                                             & Banking &  & \\
                                             & Carding &  & \\

\hline
\multirow{2}{*}{Cryptocurrencies}            & Cryptocurrencies & \multirow{2}{*}{392 (10.1\%)} & \multirow{2}{*}{4,437} & \multirow{2}{*}{4,829 (6.9\%)}\\
                                             & Crypto swapping and exchanges &  & \\

\hline
\multirow{5}{*}{Marketplaces}                & Hardware marketplaces & \multirow{5}{*}{245 (6.3\%)} & \multirow{5}{*}{1,614} & \multirow{5}{*}{1,859 (2.7\%)}\\
                                             & Mobile and device marketplaces &  & \\
                                             & Drug marketplaces &  & \\
                                             & Firearm marketplaces &  & \\
                                             & Generic markets &  & \\

\hline
\multirow{5}{*}{Media, forums and personal websites} & Personal websites and blogs & \multirow{5}{*}{214 (5.5\%)} & \multirow{5}{*}{1,491} & \multirow{5}{*}{1,705 (2.5\%)} \\
                                                     & News and media &  & \\
                                                     & Debian community &  & \\
                                                     & Debian conferences &  & \\

\hline
\multirow{7}{*}{Navigation pages} & Javascript pages & \multirow{7}{*}{196 (5\%)} & \multirow{7}{*}{330} & \multirow{7}{*}{526 (0.8\%)}\\
                                  & Login and register pages &  & \\
                                  & DDoS protection pages &  & \\
                                  & Redirecting pages &  & \\
                                  & Error pages &  & \\
                                  & One-line pages &  & \\
                                  & 'Index of' pages &  & \\

\hline
\multirow{2}{*}{Hacking}                     & Data leaks & \multirow{2}{*}{180 (4.7\%)} & \multirow{2}{*}{332} & \multirow{2}{*}{512 (0.7\%)}\\
                                             & Hacking &  & \\

\hline
\multirow{2}{*}{Counterfeits}               & Passports and certificates & \multirow{2}{*}{72 (1.9\%)} & \multirow{2}{*}{195} & \multirow{2}{*}{267 (0.4\%)}\\
                                            & Counterfeits &  & \\

\hline
\multirow{2}{*}{Privacy-preserving services}& Image and file hosting & \multirow{2}{*}{65 (1.7\%)} & \multirow{2}{*}{79} & \multirow{2}{*}{144 (0.2\%)}\\
                                            & Privacy-preserving services &  & \\

\hline
\multirow{3}{*}{Hiring services}            & Betting services & \multirow{3}{*}{61 (1.6\%)} & \multirow{3}{*}{56} & \multirow{3}{*}{117 (0.2\%)}\\
                                            & Hitman services &  & \\
                                            & Escrow services &  & \\
\hline
\textbf{Total} & & 3,869 (100\%) & 65,630 & 69,499 (100\%)\\
\hline
\end{tabular}
\caption{Onion services distribution per high-level topic}
\label{tab:htopics}
\end{table*}

\begin{figure*}[t!]
    \centering
    \includegraphics[width=1.05\textwidth]{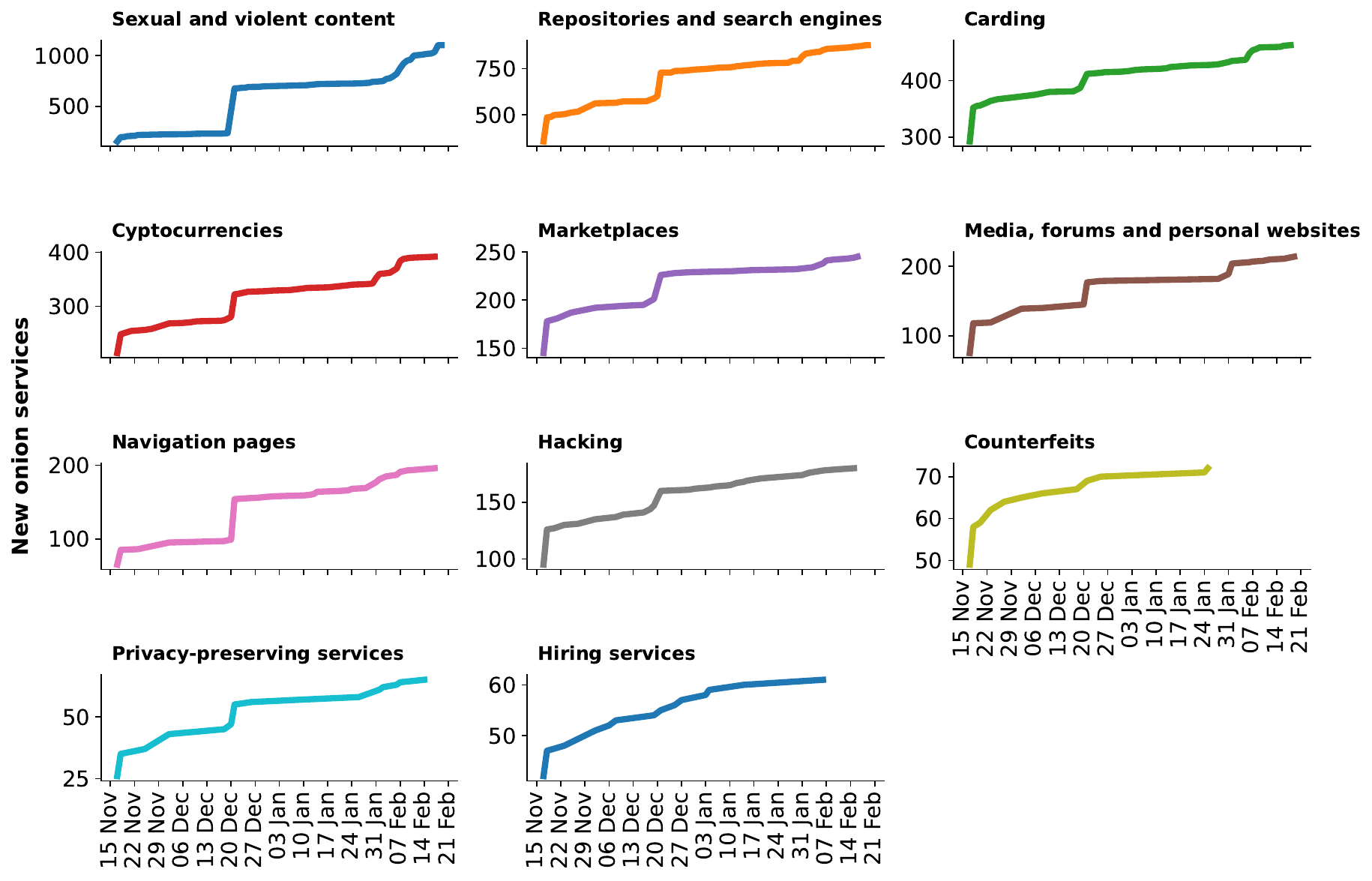}
    \caption{Accumulated identification of unique onion services per high-level topic}
    \label{fig:topicstimeline}
\end{figure*}

The architecture identifies new unique Tor sites for each theme over time. Figure~\ref{fig:topicstimeline} displays the cumulative discovery of new onion services per day, with two significant peaks on 21 December and 6-7 February, corresponding to the addition of new monitoring sources in the architecture (noted in Figure~\ref{fig:collection}). As duplicates of previously seen onion services are not included in the graph, it highlights that new Tor pages are being created daily. The trend is uniform across each topic, indicating that the Tor network gradually incorporates new and distinct content over time, making it a living space where existing onion services are not only disappearing and getting replicated.

\subsection{Exploration of duplicates}

\subsubsection{Most replicated sites}

\begin{figure*}[t!]
    \centering
    \includegraphics[width=\textwidth]{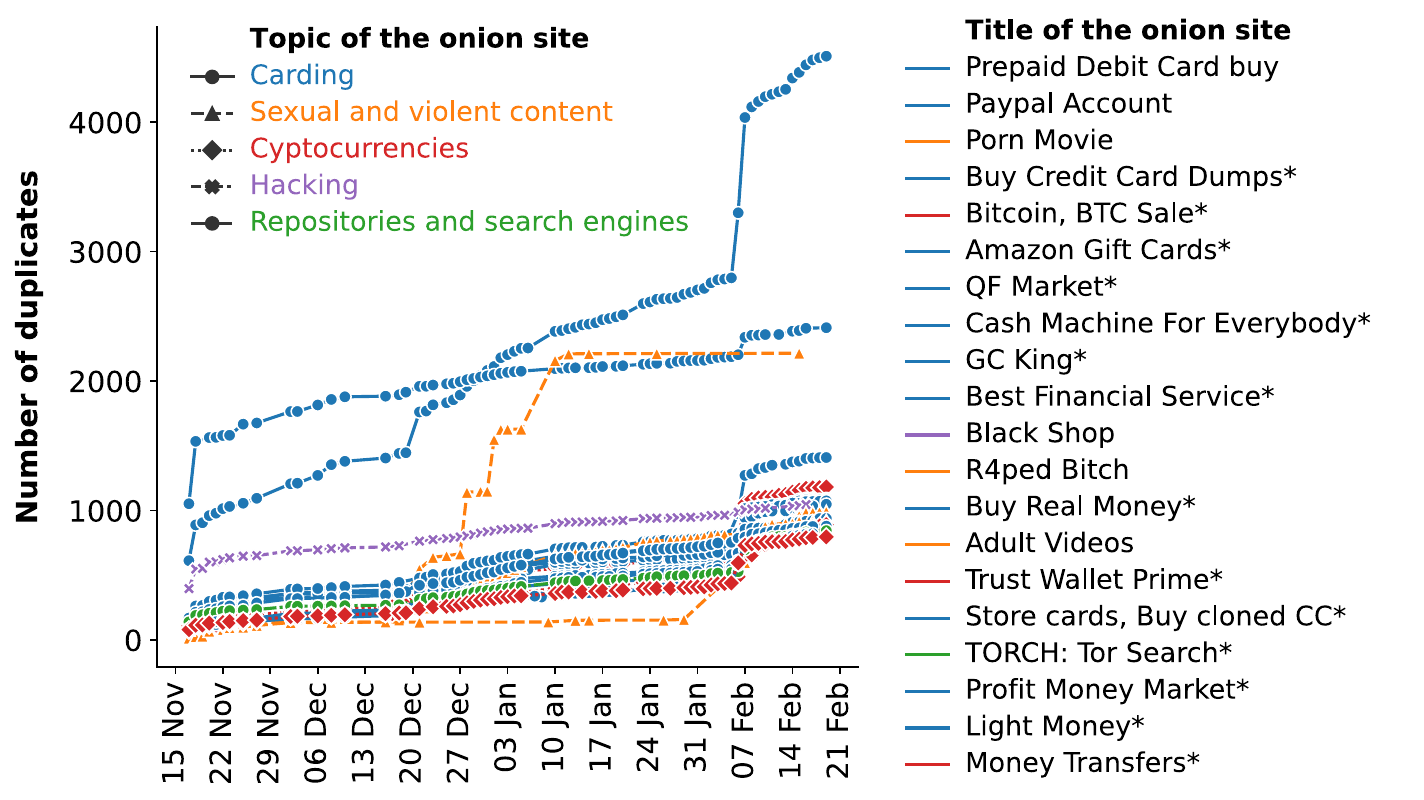}
    \caption{Duplicates of the top-20 most replicated onion services over time (asterisks indicate onions with similar trend)}
    \label{fig:top20}
\end{figure*}

Based on the disproportionate amount of duplicates discovered by the architecture, we study which particular onion services were the most replicated. Figure~\ref{fig:top20} shows the duplicates over time of the top-20 most replicated sites (along with the page title and inferred topic), which sums 26,042 instances out of 69,499 Tor sites (37.47\%).

In this ranking, 12 are related to carding, 3 with sexual and violent content, 3 with cryptocurrencies, 1 with hacking and 1 with repositories and search engines. The ``\textit{Prepaid Debit Card buy}" has more than 4,000 replicas, the ``\textit{Paypal Account}" and ``\textit{Porn Movie}" pages have more than 2,000 replicas, and the rest ranges between 800 and 1000 replicas.

At the bottom of the line plot, a set of sites share the same temporal pattern of duplicate generation, \change{suspicious phenomena that attract}{a suspicious phenomenon that attracts} the authors' attention for further inspection.

\subsubsection{Duplicated sites with similar replication ratio}

\begin{figure*}[t!]
    \centering
    \includegraphics[width=1.03\textwidth]{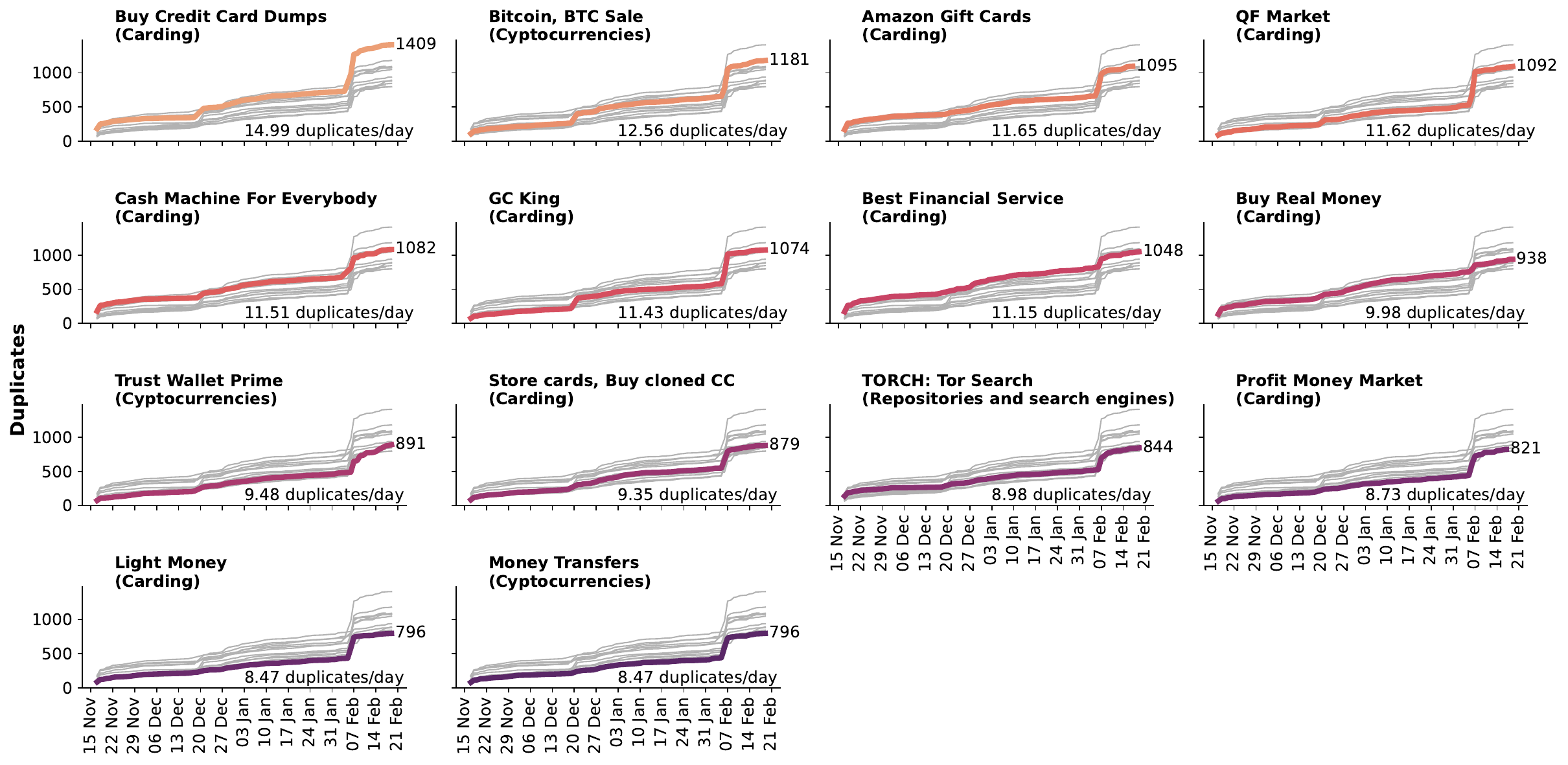}
    \caption{Duplicates of 14 onion services with a similar replication pattern.}
    \label{fig:suspicious}
\end{figure*}

Focusing on websites with similar replication trends, Figure~\ref{fig:suspicious} details the detection of duplicate pages per onion site over time. Each unique 14 sites were manually verified and found to have completely different content. However, except for the Torch repository, all sites were related to topics such as credit cards, bitcoin, gift cards, cryptocurrencies, wallets, and money, and classified by the solution under the categories of carding and cryptocurrencies.

The ratios of duplicate pages per day were very close, where the main difference lies in the initial offset caused by the number of onions detected on the first day. On average, seven pages generated between 8 and 9 duplicates per day, five pages created 11 replicas per day, one mirrored 12 times per day, and the first one was replicated 15 times per day.

Although we have no more evidence, these figures are highly coincidental, and we hypothesize that a common actor coordinates these different pages (and duplicates). Specifically, the replication may not be intended to maximize the availability or protect against DDoS attacks. Instead, it could be a massive phishing campaign through different types of sites that contain fraud cryptocurrency addresses as attack vectors, increasing the attack surface by replicating them across the Tor network. For example, ``\textit{Light Money}" and ``\textit{Money Transfers}" are two of these services that belong to different topics but coincide in the number of duplicates (796) and replication ratio (8.47 duplicates/day), showing an identical trend throughout the collection window. Moreover, the Torch repositories could be deployed to reference the illegitimate pages and amplify the threat landscape.

Generally, phishing is common and effective on the Tor network due to the anonymity provided, being exploited by cybercriminals that evade detection through fake pages that can not be authenticated.

\section{Discussion}

The literature presents several approaches for analyzing Tor sites but needs to be more in adopting modern data ingestion, processing, and storage frameworks. Solutions are limited in scalability and efficacy, leading to a shortage of collections from 529 to 10,163 sites. In order to address this gap, this paper proposes an Open Source Big Data architecture for identifying and categorizing onion services in near real-time, representing unprecedented performance in Tor monitoring with 72K analyzed onion services. 

Kubernetes is a key technological enabler that facilitates a direct deployment of Kafka or integration of native complements such as MinIO and Kubeflow. Additionally, the resources of the computing nodes can be managed efficiently. In this sense, it guarantees that the continuous monitoring part is always active to detect and download the content of the onions. On the other hand, the batch processing task is punctual and dynamically launched in a finite way to characterize the sample collected every day, having most of the time those resources available in the cluster. Otherwise, the latter would be exclusively dedicated to processing and machine learning processes in streaming. 

However, developing an end-to-end pipeline from scratch for community-driven projects posed significant challenges due to gaps and technical issues. For instance, modifying low-level components like the MongoDB source connector for Kafka or Crawlab spiders was complex. Additionally, while the orchestration and deployment of microservices are theoretically simple in virtualized frameworks, integrating and communicating all the distributed elements into a common logic is really hard. Therefore, from September 1 to November 17, we spent more than two months to have the architecture up and running without inconsistencies and errors through GitLab CI/CD operations within the cluster. The result is a sophisticated platform, but not overly romanticized against monolithic architectures that can be much easier to develop for most use cases. Despite these obstacles, adopting DevOps (GitLab CI/CD with Kubernetes) and MLOps (Kubeflow) allowed us to define operations by software, integrating operation and development into one workflow.

Regarding the challenges of the Tor use case, the volatility and short life of onion services make it difficult to implement tools for monitoring the Tor network landscape. In the related works, the discovered onion services showed a low availability (from 3\% to 35\%), having a big discrepancy between the addresses identified and those active and analyzed. This paper addresses these coverage and access problems with early identification under continuous monitoring. By adopting the strategy in the pipeline, the categorization is executed before onion services cease to exist, having reached 72,045 onion sites out of 80,330 (90\% of availability), the second most representative research of the literature~\cite{javierFGCS} (after 82,145 active sites considered by~\cite{15}).

Another critical aspect faced is extracting meaningful and valuable text from raw HTML pages containing too much noise. Particularly, dark web sites are not well-formed and break the HTML best practices, which is different from standard surface sites. While the study of surface webpages is widely more common in the literature, dark web research is a bit retarded in the application of modern NLP strategies. Downloading UTF-8, parsing with BautifulSoup, extracting relevant text with Trafilatura and application of regular expressions to remove noise (addresses, links, weird strings and characters, etc.)  was hard in order to get a good performance in preprocessing messy dark HTML files.

Some studies demonstrate a substantial prevalence of duplicated dark web sites, from 20\% to 95\%, yet the majority of analytical experiments fail to account for this phenomenon, leading to an overload of processing the same content multiple times, but also generating distorted research results and distributions. In order to address this limitation, our proposed framework integrates content similarity with Jaccard distance to analyze only new never-seen documents, and MinHash LSH to increase scalability due to the reduction of document comparisons to perform. In our case, we found 93.9\% of repeated content, reducing unnecessary processing overhead and being in line with literature results.

Regarding categorization, related works rely on manual, keyword, or probabilistic-based models, with a noticeable absence of modern topic extraction or document classification based on neural networks or embeddings. Additionally, the sample sizes used in previous classifications, from 445 to 6,227 dark sites, may not be sufficient to capture the topic diversity of the Tor network. In this work, we apply BERTopic to analyze 72,045 onion services, marking the largest categorization effort among the reviewed works. This framework has shown promising results in various domains~\cite{Hanley_Kumar_Durumeric_2023,egger2022topic}, and our study serves as another example.

It is worth noting that we identified several different types of navigation pages, such as error, index-of, DDoS protection, register and logging, and redirecting, and javascript. Other findings include file hosting, privacy services, Debian communities, and Debian conference pages. In terms of defense, some were directly related to cyber warfare (hacking), data leaks, propaganda (news and media), weapons trading (firearms), or organized communities (forums). However, considering the high percentage of mirrored sites within the dark web, experimental results indicate a 95\% probability of encountering services or repositories associated with illicit trading, such as sexual and violent content, cards, cryptocurrencies, or marketplaces. Conversely, potentially ethical resources, such as forums, websites, media, or privacy services, are limited to approximately 2.5\%.

Finally, one might argue that dark web monitoring does not fit squarely within the Big Data paradigm, given the existence of around 800 thousand onions simultaneously, according to Tor metrics\footnote{https://metrics.torproject.org/hidserv-dir-v3-onions-seen.html}. While this scenario may seem limited due to the slow influx of new onions daily, the modular and virtualized \change{architectures allow}{architecture allows} for deploying more advanced Tor use cases. Examples include daily downloading and categorizing all onion samples to track changes or extending real-time ingestion to new darknets like I2P, Freenet, ZeroNet, or even the surface web. In our project, the design is sufficiently generic and extensible to support various types of workloads, from constrained use cases for easy deployment and management of the application to high-demanding scenarios requiring scalability, robustness, and adaptability.

\section{Conclusion}
In this paper, we propose an architecture for continuously obtaining and analyzing Tor onion services detected in diverse data sources, including threat intelligence, code repositories, Web-Tor gateways, and Tor repositories. We include a near real-time ingestion with a crawler to visit the data sources and use Kafka to coordinate the download of HTML pages by a set of downloaders. At the end of the day, a batch processing pipeline based on Kubeflow preprocesses the HTML content, deduplicates with MinHash LSH, and extracts languages with fasttext while inferring topics with BERTopic. The architecture is deployed under DevOps (GitLab CI/CD) and MLOps (Kubeflow) paradigms, easily monitored and configured with the Kubernetes Dashboard (data engineering), Crawlab user-friendly framework (data ingestion) and Kubeflow interface (data processing).

The solution has been running for 93 days, categorizing 72,045 onion services out of 80,049 identified. In this sense, early identification of sites using the crawler's scheduled spiders is highly effective, enabling the solution to connect and characterize 90\% of the Tor sites. This is a significant improvement from the volatility ratios reported in related works. The architecture identified 56,674 (78.7\%) exact duplicates (hash matching) and 10,640 (14.8\%) near duplicates (MinHash LSH), having only 4,390 (6.1\%)  unique sites. This disproportional amount of repeated content is even higher than in other literature reviews and exposes most Tor-based studies that do not consider this phenomenon in their interpretations and results. 

The predominant language is English, with 88.09\% in the unique sample and 96,9\% in the overall dataset with duplicates. The BERTopic methodology is applied to this subset of documents, returning 35 low-level topics, which the authors manually merged \change{to}{into} 11 considering the cosine distance between topic embeddings: sexual and violent content; repositories and search engines; carding; cryptocurrencies; marketplaces; media, forums and personal websites; navigation pages; hacking; counterfeits; privacy-preserving services; and hiring services (in order of prevalence). Over the days, the appearance of onions in each topic has a similar pattern and does not show significant peaks or anomalies.

In the exploration of duplicates, we specifically studied the clone distribution of the 20 most replicated onion services (37.47\% of the dataset), discovering a subset of 14 Tor sites of different nature that maintain an almost identical daily appearance pattern. While the detailed study of these remains as future work, it seems to us that these duplicates of different onion services would not be focused on mirroring or protection against DDoS attacks, but part of a coordinated phishing network that automatically multiplies its exposure on the dark web through different types of content and appearance.

\section{Future work}

Our proposed architecture, in its current state, has proven fruitful in aggregating and characterizing Tor onion services. Nonetheless, several promising directions for future work remain to be explored. 

A prevalent observation from our study is the existence of a substantial number of duplicated dark websites. The development of advanced techniques to discriminate between genuine and phishing sites is necessary to be integrated into the ML pipeline. Such an advancement would not only enhance data accuracy but also fortify subsequent analyses' reliability. Additionally, potential advancements could stem from further enhancing the categorization process. The exploration of alternative topic modeling algorithms and comparison of their efficacy with BERTopic could yield interesting insights. Additionally, considering the global nature of Tor networks, broadening the scope of our analysis to include non-English onion services could shed light on region-specific behaviors and trends, offering a more holistic understanding of the Tor landscape.

Beyond Tor services, exploring other anonymous spaces, including a wider variety of darknets like I2P, Freenet or ZeroNet, promises a more encompassing interpretation of the dark web sphere. Concurrently, our architecture's scalability and robustness should undergo additional rigorous testing with larger data volumes and extended operational periods. These steps would accurately assess our architecture's ability to handle larger datasets and lead to the developing of more resilient systems.

Finally, the proposed platform is independent of its current Tor-based application. We foresee its potential in exploring alternative Open Source Intelligence (OSINT) use cases, from tracking cyber threats to monitoring illicit activities within the darknet. In particular, the platform could be extended with a live dashboard exhibiting real-time updates and thorough statistical analyses would significantly enhance the interpretability and accessibility of the data. Such a tool could provide invaluable insights for researchers and law enforcement agencies.

\section*{Acknowledgments}
This study was partially funded by \textit{(a)}
the Spanish Government with the FPU18/00304 contract and EST22/00738 mobility grant, and by \textit{(b)} the strategic project ``Development of Professionals and Researchers in Cybersecurity, Cyberdefense and Data Science (CDL-TALENTUM)" from the Spanish National Institute of Cybersecurity (INCIBE) and by the Recovery, Transformation and Resilience Plan, Next Generation EU. We would also like to thank all the colleagues from the Cyber-Defence campus office in Lausanne for their kind support during this internship project.

\bibliographystyle{IEEEtran}
\bibliography{main}

\end{document}